\documentclass[final,5p,times]{elsarticle}

\usepackage{amsfonts}
\usepackage{amssymb,bbold}
\usepackage{amsmath}
\usepackage{float}
\usepackage{natbib}
\usepackage{graphicx}
\usepackage{hyperref}

\hypersetup{
  colorlinks=true,linkcolor=blue,citecolor=blue,
  filecolor=blue,urlcolor=blue,breaklinks=true
}

\biboptions{sort&compress}

\journal{Physics Letters B}

\begin{document}

\begin{frontmatter}

\title{The 2D $\kappa$-Dirac oscillator}

\author[uepg]{Fabiano M. Andrade}
\ead{fmandrade@uepg.br}
\author[ufma]{Edilberto O. Silva}
\ead{edilbertoo@gmail.com}

\address[uepg]{
  Departamento de Matem\'{a}tica e Estat\'{i}stica,
  Universidade Estadual de Ponta Grossa,
  84030-900 Ponta Grossa-PR, Brazil
}
\address[ufma]{
  Departamento de F\'{i}sica,
  Universidade Federal do Maranh\~{a}o,
  Campus Universit\'{a}rio do Bacanga,
  65085-580 S\~{a}o Lu\'{i}s-MA, Brazil
}

\begin{abstract}
In this Letter, 2D Dirac oscillator in the quantum deformed framework
generated by the $\kappa$-Poincar\'{e}-Hopf algebra is considered.
The problem is formulated using the $\kappa$-deformed Dirac equation.
The resulting theory reveals that the energies and wave functions of the
oscillator are modified by the deformation parameter.
\end{abstract}


\end{frontmatter}

\date{\today}

\section{Introduction}
\label{sec:introduction}

The Dirac oscillator, established in 1989 by Moshinsky and Szczepaniak
\cite{JPA.1989.22.817,Book.Moshinsky.1996}, is considered a natural
framework to access the relativistic quantum properties of quantum
harmonic oscillator-like systems.
This model has inspired a great deal of investigations in recent years.
These studies have allowed the exploration of new models in theoretical
and experimental physics.
In the context of recent investigations, the interest in this issue
appears, for example, in quantum optics
\cite{PRA.2007.76.041801,PRA.2008.77.033832}, deformed Kempf algebra
\cite{SR.2013.3.3221}, graphene physics \cite{arXiv:1311.2021},
noncommutative space
\cite{JMP.2014.55.032105,IJMPA.2011.26.4991,IJTP.2012.51.2143}, quantum
phase transition \cite{PRA.2008.77.063815,arXiv:1312.5251} and
topological defects \cite{AP.2013.336.489,PRA.2011.84.32109}.
Among several recent contributions on the Dirac oscillator, we refer to
its first experimental verification \cite{PRL.2013.111.170405}.
For a more detailed approach of the Dirac oscillator see Refs.
\cite{JPA.1997.30.2585,AIPCP.2011.1334.249,Book.1998.Strange,
PRA.1994.49.586,EPJB.2001.22.31,MPLA.2004.19.2147,arXiv:1403.4113}.

The dynamics of the Dirac oscillator is governed by the Dirac equation
with the nonminimal prescription
\begin{equation}\label{eq:prescription}
  \mathbf{p}\rightarrow \mathbf{p}-im\omega\tilde{\beta}\mathbf{r},
\end{equation}
where $\mathbf{p}$ is the momentum operator, $m$ is the mass, $\omega$
the frequency of the oscillator and $\mathbf{r}$  is the position
vector.
In the same context, another usual framework where one can study the
dynamics of the Dirac oscillator is that in connection with the theory
of quantum deformations.
These quantum deformations are realized based on the
$\kappa$-Poincar\'{e}-Hopf algebra
\cite{PLB.1991.264.331,PLB.1992.293.344,PLB.1994.329.189,
PLB.1994.334.348}
and has direct implications on the quantum dynamics of relativistic and
nonrelativistic quantum systems.
The deformation parameter $\kappa$ appearing in the theory is usually
interpreted as being the Planck mass $m_{P}$ \cite{PLB.2012.711.122}.
Some important contributions on $\kappa$-deformation have been studied
in Refs.
\cite{AoP.1995.243.90,CQG.2010.27.025012,
NPB.2001.102-103.161,EPJC.2003.31.129,PLB.2002.529.256,
PRD.2011.84.085020,JHEP.2011.1112.080,EPJC.2013.73.2472,
PRD.2009.79.045012,EPJC.2006.47.531,EPJC.2008.53.295,
PRD.2013.87.125009,PRD.2012.85.045029,PRD.2009.80.025014}.

The physical properties of $\kappa$-deformed relativistic quantum
systems can be accessed by solving the $\kappa$-deformed Dirac equation
\cite{PLB.1993.302.419,PLB.1993.318.613}.
Recently, some studies involving $\kappa$-deformation have also been
reported with great interest.
Some theoretical contributions in this
context can be found, for example, in Refs.
\cite{PLB.2013.719.467,PLB.1994.339.87,PRD.2007.76.125005,
PLB.1995.359.339,MPLA.1995.10.1969}.
The 3D Dirac oscillator has been discussed in connection with the theory
of quantum deformations in Ref. \cite{PLB.2014.731.327}.
However, it is well known that the 2D Dirac oscillator exhibits a
dynamics completely different from that of 3D one.
In this context, the main goal of this Letter is study the
dynamics of the 2D Dirac oscillator in the quantum deformed framework
generated by the $\kappa$-Poincar\'{e}-Hopf algebra and after compare
with the usual (undeformed) 2D Dirac oscillator.

This Letter is organized as follow.
In Section \ref{sec:2ddiraco}, we revise the 2D Dirac oscillator and
determine the energy levels and wave functions.
In Section \ref{sec:kappadirac}, the 2D Dirac oscillator in the
framework of the quantum deformation is discussed.
A brief conclusion is outlined in Section \ref{sec:conclusion}.

\section{The 2D Dirac oscillator}
\label{sec:2ddiraco}

In this section, we briefly discuss the usual 2D Dirac oscillator
for later comparison with the deformed one.
One begins by writing the Dirac equation for the four-component spinor 
$\Psi$
\begin{equation}\label{eq:dirac}
  \left(
    \tilde{\beta}\tilde{\boldsymbol{\gamma}} \cdot
    \mathbf{p}+\tilde{\beta} m
  \right)\Psi=E\Psi.
\end{equation}
The 2D Dirac oscillator is obtained throught the nonmininal prescription
in Eq. \eqref{eq:prescription} where $\mathbf{r}=(x,y)$ is the position
vector.
We shall now make use of the underlying symmetry of the system to reduce
the four-component Dirac equation to a two-component spinor equation.
We use the following representation for the $\tilde{\gamma}$ matrices
\cite{PRD.1978.18.2932,NPB.1988.307.909,PRL.1989.62.1071}
\begin{align}
  \tilde{\beta}=\tilde{\gamma}_{0}=
  \left(
    \begin{array}{cc}
      \sigma_{z} & 0 \\
      0          & -\sigma_{z}
    \end{array}
  \right),
   & \qquad
  \tilde{\gamma}_{1}=
  \left(
    \begin{array}{cc}
      i\sigma_{y} & 0 \\
      0          & -i\sigma_{y}
    \end{array}
  \right),\\
  \tilde{\gamma}_{2}=
  \left(
    \begin{array}{cc}
     - i\sigma_{x} & 0 \\
      0          & i\sigma_{x}
    \end{array}
  \right),
  & \qquad
  \tilde{\gamma}_{3}=
  \left(
    \begin{array}{cc}
      0          & 1 \\
      -1          & 0
    \end{array}
  \right),
\end{align}
where $\sigma_{i}$ are the Pauli matrices.

\begin{figure}
  \centering
  \includegraphics*[width=\columnwidth]{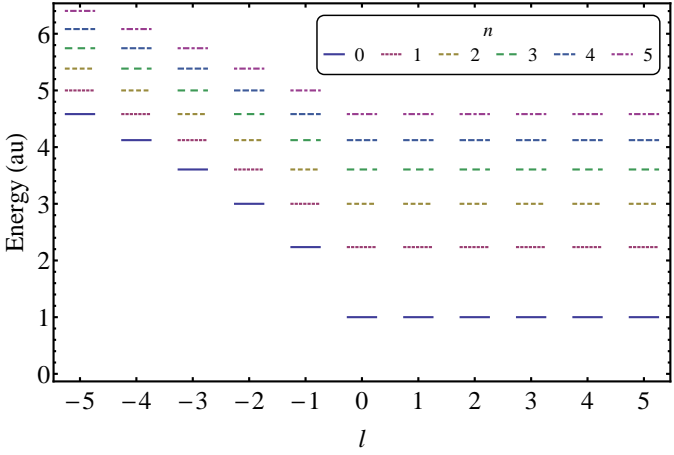}
  \caption{ \label{fig:fig1}
    (Color online) The positive energy spectrum,
    Eq. \eqref{eq:energy2ddo}, for the 2D Dirac oscillator for different
    values of $n$ and $l$ with $m=\omega=1$ and $s=1$.
    Notice that levels with quantum numbers $n \pm q$ have the same
    energy as levels with $l \pm q$, with $q$ an integer.}
\end{figure}

The 2D Dirac oscillator being independent of the $z$ direction, allow us
decouple the usual four-component Dirac equation \eqref{eq:dirac} into
two two-component equations
\begin{equation}\label{eq:dirac_oscillator}
  H_{0}\psi=\left[
    \beta \gamma \cdot
    (\mathbf{p}-i m\omega \beta\mathbf{r})+\beta m
  \right]\psi=E\psi,
\end{equation}
where $\psi$ is a two-component spinor and the three-dimensional
$\gamma$ matrices are \cite{PRL.1990.64.2347,IJMPA.1991.6.3119}
\begin{equation}
  \beta = \gamma_{0} = \sigma_{z}, \qquad
  \gamma_{1}=i\sigma_{y}, \qquad
  \gamma_{2}=-is\sigma_{x},
\end{equation}
where the parameter $s$, which is twice the spin value, can be
introduced to characterize the two spin states \cite{EPJC.2014.74.2708},
with $s=+1$ for spin ``up'' and $s=-1$ for spin ``down''.
By squaring Eq. \eqref{eq:dirac_oscillator}, one obtains
\begin{equation}\label{eq:2ddiracoscillator}
  \left[
    p^2+m^2\omega^2r^2-2m\omega(\sigma^{3}+s L_{3})
  \right]\psi=
  (E^2-m^{2})\psi.
\end{equation}
If one adopts the following decomposition
\begin{equation}\label{eq:ansatz}
  \psi=
  \left(
    \begin{array}{c}
      \psi_{1} \\
      \psi_{2}
    \end{array}
  \right)=
  \left(
    \begin{array}{c}
      f(r)\;e^{i l \phi} \\
      i g(r)\;e^{i(l+s)\phi}
    \end{array}
  \right),
\end{equation}
for the spinor, from Eq. \eqref{eq:2ddiracoscillator}, it is possible to
obtain the energy spectrum \cite{arXiv:1403.4113}:
\begin{equation}\label{eq:energy2ddo}
E=\pm\sqrt{m^{2}+2m\omega\left(2n+|l|-sl\right)},
\end{equation}
and unnormalized wave functions
\begin{equation}\label{eq:eigenfunction_2d_dirac}
  f(\rho)
  =\rho^{(|l|+1)/2} e^{-\rho/2}\;
  M\left(-n,1+|l|,\rho\right).
\end{equation}
Here, $n=0,1,2,\ldots$, $l=0,\pm 1, \pm 2,\ldots$, $\rho=m\omega r^2$
and $M(z)$ is the confluent hypergeometric function of the first kind
\cite{Book.1972.Abramowitz}.
It should be noted that the spectrum is spin dependent, and for $sl>0$
the energy eigenvalues are independent of $l$ as depicted in the
Fig. \ref{fig:fig1}.

\section{The 2D $\kappa$-Dirac Oscillator}
\label{sec:kappadirac}

In this section, we address the 2D Dirac oscillator in the framework of
$\kappa$-Poincar\'{e}-Hopf algebra.
We begin with the $\kappa$-Dirac equation defined in
\cite{PLB.1993.302.419,PLB.1993.318.613,PLB.1995.359.339} when the third
spatial coordinate is absent.
So, using the same reasoning of the previous section we have
\begin{equation}
  \left\{
    \gamma_{0} P_{0}-\gamma_{i}P_{i}+
    \frac{\varepsilon}{2}
    \left[
      \gamma_{0}(P_{0}^{2}-P_{i}P_{i})-m P_{0}
    \right]
  \right\}\psi=m\psi,
\end{equation}
with $i=1,2$.
Identifying $P_{0}=H=E$ and $P_{i}=\pi_{i}=p_{i}-im\omega\beta r_{i}$,
we have
\begin{equation}\label{eq:def_dirac}
  H\psi=
  \left[
    H_{0}-\frac{\varepsilon}{2}(H^{2}-\pi_{i}\pi_{i}-m\beta H)
  \right]
  \psi=E\psi,
\end{equation}
By iterating Eq. \eqref{eq:def_dirac}, we have up to
$\mathcal{O}(\varepsilon)$ ($\varepsilon^{2}\approx 0$)
\begin{equation}\label{eq:def_dirac_oe}
  H\psi=
  \left[
    H_{0}-\frac{\varepsilon}{2}(H_{0}^{2}-\pi_{i}\pi_{i}-m\beta H_{0})
  \right]
  \psi=E\psi,
\end{equation}
with $H_{0}$ given in Eq. \eqref{eq:dirac_oscillator}.
By using the same representation of the previous section for the
$\gamma$ matrices, Eq. \eqref{eq:def_dirac_oe} assumes the form
\begin{multline}\label{eq:ddiracocart}
  \left(1+\frac{m\varepsilon}{2}\sigma_{z}\right)
  (\sigma_{x}\pi_{x}+s\sigma_{y}\pi_{y}+m\sigma_{z})\psi=\\
  \left\{
    E+\varepsilon
    \left[
      m^2\omega^2r^2-m\omega sL_{z}+
      im\omega\sigma_{z}(\mathbf{r}\cdot\mathbf{p})
    \right]
  \right\}\psi.
\end{multline}
Equation \eqref{eq:ddiracocart}, in polar coordinates $(r,\phi)$, reads
\begin{multline}\label{eq:ddiracopolar}
  e^{is\sigma_{z}\phi}
  \left(1+\frac{m\varepsilon}{2}\sigma_{z}\right)
  \left[
    \sigma_{x}\partial_{r}+
    \sigma_{y}\left(\frac{s}{r}\partial_{\phi}-im\omega r \right)
  \right]\psi=\\
  i\left[
    E-m\sigma_{z}+\varepsilon
    \left(
      m^{2}\omega^{2}r^{2}+im\omega s\partial_{\phi}+
      m\omega\sigma_{z}r\partial_{r}
    \right)
  \right]\psi.
\end{multline}

Our task now is solve Eq. \eqref{eq:ddiracopolar}.
As the deformation does not break the angular symmetry, then we can use a
similar ansatz as for the usual (undeformed) case
\begin{equation}\label{eq:ansatzeps}
  \psi=
   \left(
    \begin{array}{c}
      f_{\varepsilon}(r)\;e^{i l \phi} \\
      i g_{\varepsilon}(r)\;e^{i(l+s)\phi}
    \end{array}
  \right),
\end{equation}
but now with the radial part labeled by the deformation parameter.
So, by using the ansatz \eqref{eq:ansatzeps} into
Eq. \eqref{eq:ddiracopolar}, we find a set of two coupled radial
differential equations of first order
\begin{subequations}
\begin{multline}\label{eq:ddiracopolarcomp1}
\left(1+\frac{m\varepsilon}{2}\right)
  \left[
    \frac{d}{dr}+
    \frac{s(l+s)}{r}-m\omega r
  \right]g_{\varepsilon}(r)= (E-m)f_{\varepsilon}(r)\\
  +    \varepsilon
    \left(
      m^{2}\omega^{2}r^{2}-m\omega s l +
      m\omega r \frac{d}{dr}
    \right)f_{\varepsilon}(r),
\end{multline}
\begin{multline}\label{eq:ddiracopolarcomp2}
\left(1-\frac{m\varepsilon}{2}\right)
  \left[
    \frac{d}{dr}-
    \frac{sl}{r}+m\omega r
  \right]f_{\varepsilon}(r)= -(E+m)g_{\varepsilon}(r)\\
  -\varepsilon
    \left(
      m^{2}\omega^{2}r^{2}-m\omega s(l+s)-
      m\omega\rho \frac{d}{dr}
    \right)g_{\varepsilon}(r).
\end{multline}
\end{subequations}
The above system of equations can be decoupled yielding a single second
order differential equation for $f_{\varepsilon}(r)$,
\begin{multline} \label{eq:edof}
  -f_{\varepsilon}''(r)-
  \left(
    \frac{1}{r}+2m^{2}\varepsilon \omega r
  \right)f_{\varepsilon}'(r)\\
  + \left[
    \frac{l^{2}}{r^{2}}+
    \left(1-2 E\varepsilon \right) m^{2}\omega^{2}r^{2}-
    k_{\varepsilon}^{2}
  \right] f_{\varepsilon}(r) =0,
\end{multline}
where
\begin{equation}\label{eq:mue}
  k_{\varepsilon}^{2}=E^2-m^2+
    2m\omega\left[(sl+1)(1-\varepsilon E)+m\varepsilon\right].
\end{equation}
A similar equation for $g_{\varepsilon}$ there exists.
The regular solution for Eq. \eqref{eq:edof} is
\begin{equation}\label{eq:def_eigen}
    f_{\varepsilon}(\rho)
  =(\lambda_{\varepsilon}\;\rho)^{(|l|-1)/2}
  e^{-(m\varepsilon+\lambda_{\varepsilon})\rho/2}
  M\left(
    d_{\varepsilon},1+|l|,\lambda_{\varepsilon}\;\rho
  \right),
\end{equation}
where
\begin{equation}
\lambda_{\varepsilon}=1-E\varepsilon,
\end{equation}
and
\begin{equation}
  d_{\varepsilon}=\frac{1+|l|}{2}+
  \frac{2m^2\omega \varepsilon-k_{\varepsilon}^2}
  {4 \gamma \lambda_{\varepsilon}}.
\end{equation}

The deformed spectrum are obtained by establishing as convergence
criterion the condition $d_{\varepsilon}=-n$.
In this manner, the deformed energy spectrum is given by
\begin{equation}
  \label{eq:energy_def}
  E^{2}-m^{2}=2m\omega(2n+|l|-sl)\lambda_{\varepsilon}.
\end{equation}
Thus, solving Eq. \eqref{eq:energy_def} for $E$, the deformed energy
levels are explicitly given by
\begin{equation}\label{eq:energy_def_ex}
  E=  \pm \sqrt{m^2+2m\omega
    \left(
      2n+|l|-sl
  \right)}-m\varepsilon\omega(2n+|l|-sl),
\end{equation}
and its unnormalized wave functions are of the form
\begin{equation}\label{eq:def_eigen_def}
    f_{\varepsilon}(\rho)
  =(\lambda_{\varepsilon}\;\rho)^{(|l|-1)/2}
  e^{-(m\varepsilon+\lambda_{\varepsilon})\rho/2}
  M\left(
    -n,1+|l|,\lambda_{\varepsilon}\;\rho
  \right).
\end{equation}
In deriving our results we have neglected terms of $O(\varepsilon^2)$.

\begin{figure}
  \centering
  \includegraphics*[width=\columnwidth]{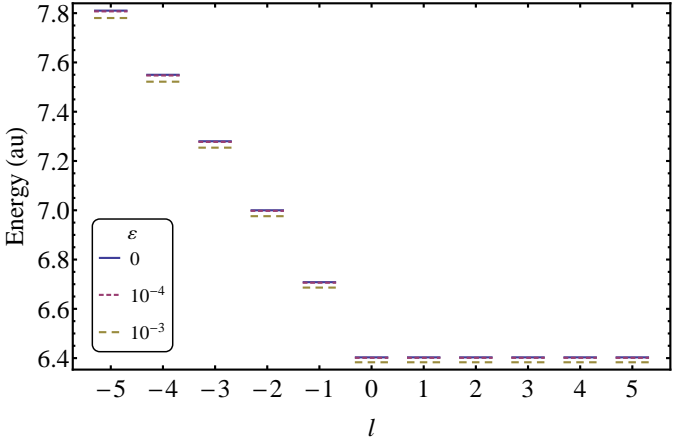}
  \caption{ \label{fig:fig2}
    (Color online) The deformed energy levels,
    Eq. \eqref{eq:energy_def_ex}, for the 2D $\kappa$-Dirac oscillator
    for $n=10$, $s=1$ and for different values of $l$.
    We use units such as $m=\omega=1$.}
\end{figure}

We can observe that the particle and antiparticle energies in the 2D
$\kappa$-Dirac oscillator are different, as a consequence of charge
conjugation symmetry breaking caused by the deformation parameter in the
same manner as observed in the three-dimensional one
\cite{PLB.2014.731.327}.
Notice that, $\varepsilon=0$, exactly conducts to the results for the
energy levels and wave functions of the previous section for the
usual (undeformed) 2D Dirac oscillator, revealing the consistency of the description
here developed.
It is worthwhile to note that the infinity degeneracy present in the
usual two-dimensional Dirac oscillator is preserved by the deformation, but
affecting the separation of the energy levels.
The distance between the adjacent energy levels decreases as the
deformation parameter increases.
In Fig. \ref{fig:fig2} is depicted the undeformed and deformed energy
levels for some values of the deformation parameter for $n=10$.
In Ref. \cite{PLB.2013.719.467}, we have determined a upper bound for
the deformation parameter.
Taking into account this upper bound, the product $m\varepsilon$ should
be smaller than $0.00116$.
In this manner, using units such as $m=1$, we must consider values for
the deformation smaller than $0.00116$.

\section{Conclusions}
\label{sec:conclusion}

In this letter, we considered the dynamics of the 2D Dirac oscillator
in the context of $\kappa$-Poincar\'{e}-Hopf algebra.
Using the fact that the deformation does not break the angular symmetry,
we have derived the $\kappa$-deformed radial differential equation whose
solution has led to the deformed energy spectrum and wave functions.
We verify that the energy spectrum and wave functions are modified by
the presence of the deformation parameter $\varepsilon$.
Using values for the deformation parameter lower than the upper
bound $0.00116$, we have examined the dependence of the energy of the
oscillator with the deformation.
The deformation parameter modifies the energy spectrum and wave
functions the Dirac oscillator, preserving the infinity degeneracy, but
affecting the distance between the adjacent energy levels.
Finally, the case $\varepsilon=0$, exactly conducts for the results for
the usual 2D Dirac oscillator.

\section{Acknowledgments}
We would like to thank Rodolfo Casana for discussions on
dimensionality of the deformed Dirac equation.
This work was supported by the
Funda\c{c}\~{a}o Arauc\'{a}ria (Grant No. 205/2013 (PPP) and
No. 484/2014 (PQ)),
and the Conselho Nacional de Desenvolvimento
Cient\'{i}fico e Tecnol\'{o}gico (Grants No. 482015/2013-6 (Universal),
No. 306068/2013-3 (PQ)) and FAPEMA (Grant No. 00845/13).

\bibliographystyle{model1a-num-names}

\end{document}